\documentclass[aps,pra,reprint,superscriptaddress]{revtex4-2}
\usepackage{booktabs}
\usepackage{amsmath}
\usepackage{amssymb}
\usepackage{hyperref}
\usepackage{multirow}
\usepackage[capitalize]{cleveref}
\usepackage[arrowdel]{physics}
\usepackage{dsfont}
\usepackage{diagbox}
\usepackage{glossaries}
\usepackage{glossaries-extra}

\makeatletter
\DeclareFontFamily{OMX}{MnSymbolE}{}
\DeclareSymbolFont{MnLargeSymbols}{OMX}{MnSymbolE}{m}{n}
\SetSymbolFont{MnLargeSymbols}{bold}{OMX}{MnSymbolE}{b}{n}
\DeclareFontShape{OMX}{MnSymbolE}{m}{n}{
	<-6>  MnSymbolE5
	<6-7>  MnSymbolE6
	<7-8>  MnSymbolE7
	<8-9>  MnSymbolE8
	<9-10> MnSymbolE9
	<10-12> MnSymbolE10
	<12->   MnSymbolE12
}{}
\DeclareFontShape{OMX}{MnSymbolE}{b}{n}{
	<-6>  MnSymbolE-Bold5
	<6-7>  MnSymbolE-Bold6
	<7-8>  MnSymbolE-Bold7
	<8-9>  MnSymbolE-Bold8
	<9-10> MnSymbolE-Bold9
	<10-12> MnSymbolE-Bold10
	<12->   MnSymbolE-Bold12
}{}

\let\llangle\@undefined
\let\rrangle\@undefined
\DeclareMathDelimiter{\llangle}{\mathopen}%
{MnLargeSymbols}{'164}{MnLargeSymbols}{'164}
\DeclareMathDelimiter{\rrangle}{\mathclose}%
{MnLargeSymbols}{'171}{MnLargeSymbols}{'171}
\makeatother

\DeclareDocumentCommand\bbra{ s m t\kket s g }
{ 
	\IfBooleanTF{#3}
	{ 
		\IfBooleanTF{#1}
		{ 
			\IfNoValueTF{#5}
			{\iipp*{#2}{} \IfBooleanTF{#4}{*}{}}
			{\iipp*{#2}{#5}}
		}
		{
			\IfBooleanTF{#4}
			{ 
				\IfNoValueTF{#5}
				{\iipp{#2}{} *}
				{\iipp*{#2}{#5}}
			}
			{\iipp{#2}{\IfNoValueTF{#5}{}{#5}}} 
		}
	}
	{ 
		\IfBooleanTF{#1}
		{\vphantom{#2}\left\llangle\smash{#2}\right\rvert}
		{\left\llangle{#2}\right\rvert}
		\IfBooleanTF{#4}{*}{}
		\IfNoValueTF{#5}{}{#5}
	}
}

\DeclareDocumentCommand\kket{ s m }
{ 
	\IfBooleanTF{#1}
	{\vphantom{#2}\left\lvert\smash{#2}\right\rrangle} 
	{\left\lvert{#2}\right\rrangle} 
}

\DeclareDocumentCommand\iipp{ s m g }
{ 
	\IfBooleanTF{#1}
	{ 
		\IfNoValueTF{#3}
		{\vphantom{#2}\left\llangle\smash{#2}\middle\vert\smash{#2}\right\rrangle}
		{\vphantom{#2#3}\left\llangle\smash{#2}\middle\vert\smash{#3}\right\rrangle}
	}
	{ 
		\IfNoValueTF{#3}
		{\left\llangle{#2}\middle\vert{#2}\right\rrangle}
		{\left\llangle{#2}\middle\vert{#3}\right\rrangle}
	}
}

\DeclareDocumentCommand\oopp{ s m g }
{ 
	\IfBooleanTF{#1}
	{ 
		\IfNoValueTF{#3}
		{\vphantom{#2}\left\lvert\smash{#2}\middle\rrangle\!\middle\llangle\smash{#2}\right\rvert}
		{\vphantom{#2#3}\left\lvert\smash{#2}\middle\rrangle\!\middle\llangle\smash{#3}\right\rvert}
	}
	{ 
		\IfNoValueTF{#3}
		{\left\lvert{#2}\middle\rrangle\!\middle\llangle{#2}\right\rvert}
		{\left\lvert{#2}\middle\rrangle\!\middle\llangle{#3}\right\rvert}
	}
}

\DeclareDocumentCommand\eevv{ s s m g }
{ 
	\IfNoValueTF{#4}
	{
		\IfBooleanTF{#1}
		{\vphantom{#3}\left\llangle\smash{#3}\right\rrangle} 
		{\left\llangle{#3}\right\rrangle} 
	}
	{
		\IfBooleanTF{#1}
		{
			\IfBooleanTF{#2}
			{\left\llangle{#4}\middle\vert{#3}\middle\vert{#4}\right\rrangle} 
			{\vphantom{#3#4}\left\llangle\smash{#4}\middle\vert\smash{#3}\middle\vert\smash{#4}\right\rrangle} 
		}
		{\vphantom{#3}\left\llangle{#4}\middle\vert\smash{#3}\middle\vert{#4}\right\rrangle} 
	}
}

\DeclareDocumentCommand\mmel{ s s m m m }
{ 
	\IfBooleanTF{#1}
	{
		\IfBooleanTF{#2}
		{\left\llangle{#3}\middle\vert{#4}\middle\vert{#5}\right\rrangle} 
		{\vphantom{#3#4#5}\left\llangle\smash{#3}\middle\vert\smash{#4}\middle\vert\smash{#5}\right\rrangle} 
	}
	{\vphantom{#4}\left\llangle{#3}\middle\vert\smash{#4}\middle\vert{#5}\right\rrangle} 
}

\newabbreviation{Gls:QN}{QN}{quantum number}

\begin{document}
	
	\begin{abstract}
		We reformulate the Floquet theory for periodically driven quantum systems following a perfect analogy with the proof of Bloch theorem. We observe that the current standard method for calculating the Floquet eigenstates by the quasi-energy alone is incomplete and unstable, and pinpoint an overlooked quantum number, the average energy. This new quantum number resolves many shortcomings of the Floquet method stemming from the quasi-energy degeneracy issues, particularly in the continuum limit. Using the average energy quantum number we get properties similar to those of the static energy, including a unique lower-bounded ordering of the Floquet states, from which we define a ground state, and a variational method for calculating the Floquet states. This is a first step towards reformulating Floquet first-principles methods, that have long been thought to be incompatible due to the limitations of the quasi-energy.
	\end{abstract}
	
	\title{The missing quantum number of the Floquet states}
	
	\author{Cristian M. \surname{Le}}
	\email{cristian.le@phys.s.u-tokyo.ac.jp}
	\author{Ryosuke \surname{Akashi}}
	\affiliation{Department of Physics, University of Tokyo, Hongo, Tokyo, 113-0033, Japan}
	\author{Shinji \surname{Tsuneyuki}}
	\affiliation{Department of Physics, University of Tokyo, Hongo, Tokyo, 113-0033, Japan}
	\affiliation{ISSP, University of Tokyo, Kashiwa, Tokyo, 277-8581 Japan}
	
	\date{\today}
	\maketitle
	\section{Introduction}
	\begin{table*}
		\renewcommand{\arraystretch}{1.4}
		\begin{tabular}{|l|c|c|c|}
			\toprule
			\multirow{2}{*}{}&\multirow{2}{*}{Bloch theorem}&\multicolumn{2}{c|}{Floquet theory}\\\cline{3-4}
			&&Quasi-energy only&With average energy\\\midrule
			Eigenstates&$(u_{\va{k}n}(\va{r}),\va{k},E_{\va{k}n})$&$(\Phi_n(t),\epsilon_n)$&$(\Phi_{na}(t),\epsilon_n,\bar{E}_{na})$\\\hline
			Symmetry \glsxtrshort{Gls:QN}&$\hat{\va{k}}\ket{u_{\va{k}n}}=\va{k}\ket{u_{\va{k}n}}$&\multirow{2}{*}{$\hat{H}^F\ket{\Phi_n}=\epsilon_n\ket{\Phi_n}$}&$\hat{H}^F\ket{\Phi_{na}}=\epsilon_n\ket{\Phi_{na}}$\\\cline{1-2}\cline{4-4}
			Ordering \glsxtrshort{Gls:QN}&$\hat{H}_{\va{k}}\ket{u_{\va{k}n}}=E_{\va{k}n}\ket{u_{\va{k}n}}$&&$\hat{\bar{H}}_n\ket{\Phi_{na}}=\bar{E}_{na}\ket{\Phi_{na}}$\\\hline
			Ritz variation&$E[u]\geq E_0$&$\nexists$&$\bar{E}[\Phi]\geq\bar{E}_0$
			\\\bottomrule
		\end{tabular}
		\caption{Perfect analogy of Floquet theory and Bloch theorem\cite{Ashcroft_1976}.\label{Tab:Floq-Bloch0}}
	\end{table*}

	Periodically driven quantum systems have been gaining increasing attention, particularly due to the promise of Floquet engineering~\cite{Holthaus_2015,Oka_2019} to control material properties and achieve exotic states such as laser induced superconductivity~\cite{Fausti_2011,Dasari_2018,Takasan_2017}, or topological insulators~\cite{Fregoso_2013,Klinovaja_2016}. So far these applications have been investigated only on simple model systems due to the limitations of the current Floquet methodology. In order to extend the Floquet formalism to self-consistent first-principles calculations and bigger more complex systems, we need to reevaluate the source of these limitations.

	Since the original derivations of the Floquet formalism in quantum mechanics~\cite{Shirley_1965,Sambe_1973,Okuniewicz_1974}, the fundamental methodology of solving the time-periodic Schr\"{o}dinger equation using Floquet theory has remained unchanged, relying on the definition of the quasi-energy eigenstates defined by the eigenproblem
	\begin{equation}
		\hat{H}^F(t)\ket{\Phi_n(t)}=\epsilon_n\ket{\Phi_n(t)}, \label{Eq:Eig Floquet}
	\end{equation}
	where $\hat{H}^F,\Phi$ and $\epsilon$ denote the Floquet Hamiltonian ($\hat{H}^F(t)=\hat{H}(t)-i\hat{\partial}_t$), the time-periodic Floquet states ($\Phi(t+T)=\Phi(t)$), and the quasi-energies respectively. However, we know that there are various limitations to this definition~\cite{Hone_1997,Kohn_2001,Langemeyer_2014}: the quasi-energy ordering being meaningless, the breakdown of this eigenproblem in the continuum system, the lack of a Hilbert space truncation method, and so on. These limitations are currently the main reasons why the Floquet formalism has not been generalized to various first principles calculation methods~\cite{Maitra_2002,Maitra_2007,Kapoor_2013}.
	
	As for the similarities between Floquet and Bloch theories, these have long been known~\cite{Holthaus_2015,Dittrich_1998}, but the detailed analogy between the two theories has not been thoroughly explored. If we explore this analogy (\cref{Tab:Floq-Bloch0}), we find that the current Floquet formalism is apparently incomplete, using only one quantum number to label the eigenstates as opposed to the two in Bloch systems. It should be clarified that the quasi-energy $\epsilon$ is analogous to the crystal momentum $\va{k}$, as opposed to the common misconception that it corresponds to the static energy $E$. Additionally, the Floquet method lacks a systematic derivation of the Floquet eigenstates from the fundamental symmetry of the time-periodic Hamiltonian and discrete time translation, which is otherwise well established in the Bloch theory~\cite{Ashcroft_1976}. Here we will present this missing derivation, completing the Floquet picture and Floquet-Bloch analogy, showing that the average energy~\cite{Le_2020}, our missing \gls{Gls:QN}, is the analogue of the static energy, and as such we can expect it to fulfil similar roles.
	
	In \cref{Sec:Rederive}, we are rederiving the Floquet eigenstates from first-principles, following the same steps as those in the Bloch systems~\cite{Ashcroft_1976}. We will discuss the main differences between this method and the conventional Floquet approach in \cref{Sec:Difference}, and then the properties of the average energy as a \gls{Gls:QN} in \cref{Sec:Properties}. We will conclude with \cref{Sec:Conclusion}, where we discuss some potential applications that are to come following this reformulation.

	\section{Rederiving the Floquet eigenstates\label{Sec:Rederive}}
	First we have to bring the time-periodic problem to a similar Hilbert space as the spatially periodic system. For this we promote the time parameter $t$ to an operator defined on a Lebesgue space over the whole real domain $\mathds{E}=L^2(\mathds{R},t)$. The time-dependent Hamiltonian is then defined as a self-adjoint operator on the extended Hilbert space $\mathds{H}\otimes\mathds{E}$, where $\mathds{H}$ is the Hilbert space upon which the Hamiltonian $\hat{H}(t)$ acts on at any time $t$. In this representation, the state vectors are expressed as
	\begin{gather}
		\ket{\Psi}=\int_{-\infty}^{+\infty}\ket{\Psi(t)}\otimes\ket{t}\dd{t},\\
		\ip{\Psi}{\Psi'}=\int_{-\infty}^{+\infty}\ip{\Psi(t)}{\Psi'(t)}\dd{t},
	\end{gather}
	and the time-periodic Hamiltonian is naturally extended to the following form:
	\begin{equation}
		\hat{H}=\int_{-\infty}^{+\infty}\hat{H}(t)\otimes\op{t}\dd{t}. \label{Eq:Def H}
	\end{equation}
	
	The physicality of this Hilbert space extension is a topic of great debate~\cite{Muga_2008}, with the most prominent counterpoint being that the extended Hilbert space $\mathds{H}\otimes\mathds{E}$ is over-complete~\cite{Flugge_1958,Pauli_1980}, including unphysical states which do not satisfy the time-dependent Schr\"{o}dinger equation. We overcome this issue by focussing on the physical subspace $(\mathds{H}\otimes\mathds{E})_S$~\cite{Prvanovic_2018,Giovannetti_2015} defined as
	\begin{equation}
		\Psi\in(\mathds{H}\otimes\mathds{E})_S\quad\Leftrightarrow\quad\qty[\hat{H}-i\hat{\partial}_t]\ket{\Psi}=0, \label{Eq:Cond Phys}
	\end{equation}
	where the time-derivative operator $\hat{\partial}_t$ is expressed in the $\mathds{E}$ Hilbert space as:
	\begin{equation}
		\hat{\partial}_t=\int_{-\infty}^{+\infty}\ket{t}\partial_t\bra{t}\dd{t}.
	\end{equation}
	Defining the physical subspace as such can be intimidating, so instead we use the equivalence of the time-dependent Schr\"{o}dinger equation with the Floquet Schr\"{o}dinger equation, i.e. any eigensolution of \cref{Eq:Eig Floquet} corresponds to a physical state in $(\mathds{H}\otimes\mathds{E})_S$:
	\begin{equation}
		\ket{\Psi_{n}(t)}=e^{-i\epsilon_nt}\ket{\Phi_n(t)}. \label{Eq:PsinPhin}
	\end{equation}
	Here we are using the already established identities of the quasi-energy eigenstates in \cref{Eq:Eig Floquet}~\cite{Holthaus_2015,Sambe_1973}, but reserve ourselves from defining these as the final Floquet eigenstates. As our goal is not to deny the validity of this definition, but rather complete it, we will be assuming that we know the quasi-energy eigenstates exactly. We can then define the projection operator $\mathds{1}_S$ onto the physical subspace $(\mathds{H}\otimes\mathds{E})_S$ as
	\begin{equation}
		\mathds{1}_S=\sum_{n}\op{\Psi_{n}}, \label{Eq:Def 1S}
	\end{equation}
	where we require the orthonormality condition
	\begin{gather}
		\ket{\Psi_{n}}=\lim_{\mathcal{T}\to\infty}\frac{1}{\sqrt{2\mathcal{T}}}\int_{-\mathcal{T}}^{\mathcal{T}}e^{-i\epsilon_nt}\ket{\Phi_{n}(t)}\otimes\ket{t}\dd{t}, \label{Eq:Def Psin}\\
		\ip{\Psi_{m}}{\Psi_{n}}=\delta_{mn}.
	\end{gather}
	For simplicity we have taken the Hilbert space $\mathds{H}$ to be finite, although the generalization to the continuous space is straightforward. In order for the projection $\mathds{1}_S$ to be unitary and complete in $(\mathds{H}\otimes\mathds{E})_S$, the summation label $n$ in \cref{Eq:Def 1S} is limited to a single quasi-energy Brillouin zone ($\epsilon_n\in[0,\omega)$).
	
	Next, in the Hilbert space $\mathds{E}$ we define the time translation operator $\hat{T}$ that shifts the time parameter by a time period $T$:
	\begin{equation}
		\hat{T}=\int_{-\infty}^{+\infty}\op{t-T}{t}\dd{t}. \label{Eq:Def Tl}
	\end{equation}
	This operator trivially commutes with the time-periodic Hamiltonian $\hat{H}$ in the extended space $\mathds{H}\otimes\mathds{E}$ (\cref{Eq:Def H})
	\begin{equation}
		\comm{\hat{H}}{\hat{T}}=0,
	\end{equation}
	and it is the equivalent starting point of the Bloch theorem proof~\cite{Ashcroft_1976}. However, for our purposes, we need a similar commutation relation to hold within the physical subspace $(\mathds{H}\otimes\mathds{E})_S$. For that we project and redefine these operators on the physical subspace and confirm that the commutation relation still holds there.
	
	The projection of the Hamiltonian $\hat{H}$ on the physical subspace gives us the average energy operator $\hat{\bar{H}}$.
	\begin{gather}
		\hat{\bar{H}}=\mathds{1}_S\hat{H}\mathds{1}_S=\sum_{mn}\bar{H}_{mn}\op{\Psi_{m}}{\Psi_{n}}, \label{Eq:Def barH}\\
		\bar{H}_{mn}=\lim\limits_{\mathcal{T}\to\infty}\frac{1}{2\mathcal{T}}\int_{-\mathcal{T}}^{\mathcal{T}}\mel{\Psi_{m}(t)}{\hat{H}(t)}{\Psi_{n}(t)}\dd{t}. \label{Eq:Def Hmn}
	\end{gather}
	We refer to this operator as the average energy because its expectation value gives us the observable average energy $\bar{E}$, as defined in \cite{Le_2020}, for any normalized physical wave function in $(\mathds{H}\otimes\mathds{E})_S$:
	\begin{equation}
		\bar{E}[\Psi]=\ev{\hat{\bar{H}}}{\Psi}=\lim\limits_{\mathcal{T}\to\infty}\frac{1}{2\mathcal{T}}\int_{-\mathcal{T}}^{\mathcal{T}}\ev{\hat{H}(t)}{\Psi(t)}\dd{t}. \label{Eq:Def barEPsi}
	\end{equation}
	Using the definition of \cref{Eq:Def Psin} in \cref{Eq:Def Hmn}, we can see that the average energy operator $\hat{\bar{H}}$ is diagonal with respect to different quasi-energies, and it generally lifts the quasi-energy degeneracy~\cite{Le_2020}, so that we can simplify it to the following form:
	\begin{equation}
		\bar{H}_{mn}=\begin{cases}
			0&\mathrm{if}\;\epsilon_m\neq\epsilon_n,\\
			\displaystyle\frac{1}{T}\int_{0}^{T}\mel{\Phi_m(t)}{\hat{H}(t)}{\Phi_{n}(t)}\dd{t}&\mathrm{if}\;\epsilon_m=\epsilon_n.
		\end{cases}\label{Eq:Hmn}
	\end{equation}
	
	As for the translation operator $\hat{T}$, we can quickly see that it is diagonal with respect to the quasi-energy basis defined in \cref{Eq:Def 1S,Eq:Def Psin}:
	\begin{gather}
		\hat{T}_{S}=\mathds{1}_S\hat{T}\mathds{1}_S=\sum_{n}e^{-i\epsilon_nT}\op{\Psi_n}. \label{Eq:Def TS}
	\end{gather}
	From the decompositions and identities in \cref{Eq:Def TS,Eq:Def barH,Eq:Hmn}, we can conclude that the average energy operator $\hat{\bar{H}}$ and the time translation operator $\hat{T}_S$ commute nontrivially in the physical subspace $(\mathds{H}\otimes\mathds{E})_S$ (see \cref{App:Comm} for proof). We thus define the physical eigenstates $\Psi_{na}$ as the simultaneous eigenstates of these operators, with $n$ being the \gls{Gls:QN} of the quasi-energy $\epsilon_n$ and $a$ being the additional \gls{Gls:QN} of the average energy eigenvalue $\bar{E}_{na}$.
	\begin{gather}
		\comm{\hat{\bar{H}}}{\hat{T}_S}=0,\\
		\hat{T}_S\ket{\Psi_{na}}=e^{-i\epsilon_nT}\ket{\Psi_{na}}, \label{Eq:Phys Eig2}\\
		\hat{\bar{H}}\ket{\Psi_{na}}=\bar{E}_{na}\ket{\Psi_{na}}. \label{Eq:Phys Eig1}
	\end{gather}
	This definition of the physical eigentriplet $(\Psi_{na},e^{-i\epsilon_nT},\bar{E}_{na})$ is the first major result we want to emphasize in this work.

	Finally, we substitute the Floquet state identity in \cref{Eq:PsinPhin} for the eigenstates
	\begin{equation}
		\ket{\Psi_{na}(t)}=e^{-i\epsilon_nt}\ket{\Phi_{na}(t)}, \label{Eq:Def Psina}
	\end{equation}
	into \cref{Eq:Phys Eig1,Eq:Phys Eig2} to get the equivalent eigenproblem in the Floquet space $\mathds{H}\otimes\mathds{T}$, where $\mathds{T}=L^2([0,T],t)$ is the Fourier space,
	\begin{gather}
		e^{i\epsilon_nt}\hat{T}_Se^{-i\epsilon_nt}\ket{\Phi_{na}(t)}=e^{-i\epsilon_nT}\ket{\Phi_{na}(t)}, \label{Eq:PreFloq Eig2}\\
		e^{i\epsilon_nt}\hat{\bar{H}}e^{-i\epsilon_nt}\ket{\Phi_{na}(t)}=\bar{E}_{na}\ket{\Phi_{na}(t)}. \label{Eq:PreFloq Eig1}
	\end{gather}
	\Cref{Eq:PreFloq Eig2} can be simply exchanged for the usual quasi-energy Floquet eigenproblem \cref{Eq:Eig Floquet}, while from \cref{Eq:PreFloq Eig1}, we get the equivalent average energy operator $\hat{\bar{H}}_n$ acting on the Floquet space $\mathds{H}\otimes\mathds{T}$ (see \cref{App:Hn} for derivation)
	\begin{gather}
		\hat{\bar{H}}_n=\sum_{ij}\bar{H}_{nij}\op{\Phi_{ni}}{\Phi_{nj}}, \label{Eq:Def Hn}\\
		\bar{H}_{nij}=\frac{1}{T}\int_{0}^{T}\mel{\Phi_{ni}(t)}{\hat{H}(t)}{\Phi_{nj}(t)}\dd{t}, \label{Eq:Hnij}
	\end{gather}
	where $i,j$ are the degenerate labels spanning the quasi-energy degenerate subspace:
	\begin{equation}
		\Phi_{ni}\in(\mathds{H}\otimes\mathds{T})_{\epsilon_n}\quad\Leftrightarrow\quad\hat{H}^F\ket{\Phi_{ni}}=\epsilon_n\ket{\Phi_{ni}}. \label{Eq:Phini}
	\end{equation}
	At this point we should recall the labelling convention used here as to not create any confusion with the different labelling conventions used in Floquet systems. Here we use: $m,n$ to be the quasi-energy labels, $i,j$ to be the quasi-energy degenerate labels, and $a,b$ to be the average energy label, being a subset of the former. We will also point out that the quasi-energy $\epsilon_n$ dependence of the average energy operator $\hat{\bar{H}}_n$ is identical to the crystal momentum $\va{k}$ dependence of the effective Hamiltonian $\hat{H}_{\va{k}}$ of the Bloch systems.

	Putting it all together, we have the fundamental commutation relation
	\begin{equation}
		\comm{\hat{H}^F}{\hat{\bar{H}}_n}=0,
	\end{equation}
	from which we redefine the Floquet eigenstates $\Phi_{na}$ to be the simultaneous eigenstates of both the Floquet Hamiltonian $\hat{H}^F$ and the average energy operator $\hat{\bar{H}}_n$, having the quantum numbers of the quasi-energy $\epsilon_n$ and average energy $\bar{E}_{na}$, respectively. Notice the additional \cref{Eq:Floq Eig1}:
	\begin{align}
		\hat{H}^F\ket{\Phi_{na}}&=\epsilon_n\ket{\Phi_{na}}, \label{Eq:Floq Eig2}\\
		\hat{\bar{H}}_n\ket{\Phi_{na}}&=\bar{E}_{na}\ket{\Phi_{na}}. \label{Eq:Floq Eig1}
	\end{align}
	
	\section{Difference with conventional Floquet method\label{Sec:Difference}}
	Conceptually, the conventional Floquet method of calculating the quasi-energy eigenstates is analogous to calculating the Brillouin zone in spatially periodic Bloch systems, and the additional step proposed here in \cref{Eq:Floq Eig1} is analogous to then calculating the energy bands structure. We can imagine the significance of this step by this analogy, but to be more concrete, we will explicitly explore in this section two main consequences of this redefinition. In the next section we will further justify it by looking at the properties of the average energy.

	Firstly, it resolves the quasi-energy degeneracy. Traditional Floquet eigenstates are ill-defined within the quasi-energy degenerate subspace, so that any rotated basis set $\qty{\Phi'_{nj}}$, where
	\begin{gather}
		\ket{\Phi'_{nj}}=\sum_{i}C_{ij}\ket{\Phi_{ni}},\\
		\ip{\Phi'_{ni}}{\Phi'_{nj}}=\delta_{ij},
	\end{gather}
	is an equally valid eigenbasis
	\begin{equation}
		\hat{H}^F\ket{\Phi'_{nj}}=\epsilon_n\ket{\Phi'_{nj}}\qquad\forall\qty{\Phi'_{nj}}.
	\end{equation}

	This makes it impossible to uniquely define the quasi-energy eigenstates in the continuum limit, and numerical calculations for these states become unstable. Conventionally, one would use an adiabatic continuation or pertrubative method to resolve this ambiguity, however, the former is ill-defined in the continuum limit~\cite{Hone_1997}, and the latter is unstable against infinitesimal perturbation~\cite{Le_Inpreparation}. The average energy offers an alternative labelling method, much more efficient then the previously mentioned ones as it requires a single calculation step, and it is applicable in the continuum system. We will comment that the current semi-adiabatic semi-diabatic Floquet methods~\cite{Hone_1997,Weinberg_2017} are equivalent and more intuitive if we perform the adiabatic continuation using the average energy~\cite{Le_Inpreparation}.

	Secondly, this labelling is robust against infinitesimal perturbations. It is natural to assume that a physical system would be unaffected by infinitesimally small perturbations within acceptable measurement constraints, i.e. if we take $\hat{H}^0(t)$ to be an unperturbed Hamiltonian, and $v(t)$ to be an arbitrary infinitesimally small perturbation, we expect that:
	\begin{gather}
		\ket{\Phi_{n}}\approx\ket{\Phi^0_n},
	\end{gather}
	where the standard notations for the perturbed and unperturbed systems is implied. This assumption, however, fails around the degeneracy, where
	\begin{equation}
		\exists\;\hat{v}(t):\quad\abs{\ket{\Phi_{ni}}-\ket{\Phi^0_{nj}}}=\order{1}\quad\forall i,j.
	\end{equation}
	It is thus evident how problematic it is when we consider the continuum system, where we have infinitely dense degeneracies, and how numerical computations become unstable.

	On the other hand, if we expand the derivation given here to include the near degeneracies, the labelling of the eigenstates becomes stable again~\cite{Le_Inpreparation}, i.e.
	\begin{gather}
		\ket{\Phi_{na}}\approx\ket{\Phi^0_{na}},\quad\epsilon_n\approx\epsilon^0_n,\quad\bar{E}_{na}\approx\bar{E}^0_{na}.
	\end{gather}
	The average energy $\bar{E}$ here is redefined to be the time-average over a finite time. More discussions on this topic can be found in our previous works~\cite{Le_2020,Le_Inpreparation}.

	Outside the quasi-energy degeneracies, there are virtually no differences in the Floquet eigenstates with the conventional Floquet method. There we can consider the unique properties that the average energy bring.

	\section{Properties of the average energy\label{Sec:Properties}}
	Firstly, all of the familiar theorems related to the energy eigenstates, such as Hellmann-Feynman theorem, have analogues with respect to both the quasi-energy and the average energy, as these follow straight from the eigen-definition in \cref{Eq:Floq Eig1,Eq:Floq Eig2}. Although the quasi-energy ones were already known~\cite{Sambe_1973}, because of the infinitely dense quasi-energy degeneracy in the continuum, the usefulness of these theorems are greatly diminished. On the other hand, including the average energy, these degeneracies are lifted, and we can more naturally apply these methodologies to the continuum system.
	
	Here we want to highlight that now a Ritz variational principle is possible:
	\begin{gather}
		\bar{E}[\Phi]=\ev*{\sum_n\hat{\bar{H}}_n}{\Phi}\geq\bar{E}_0\qquad\forall\Phi\in\mathds{H}\otimes\mathds{T}, \label{Eq:Ritz0}
	\end{gather}
	where the equality sign occurs at the Floquet ground-state. The lower-boundness of $\bar{E}_0$ is guaranteed if the Hamiltonian is lower-bound at all times $t$. However, \cref{Eq:Ritz0} is impractical in this form, as it would require the prior calculation of the full quasi-energy eigenspectra through the definition of \cref{Eq:Def Hn}, and it is only given here as such for theoretical purposes. In practice we can exchange it with a more calculable effective average energy functional $\bar{\mathcal{E}}[\Phi]$~\cite{Le_Inpreparation} (see \cref{App:Equiv} for proof) defined as
	\begin{equation}
		\bar{\mathcal{E}}[\Phi]=\frac{1}{T}\int_{0}^{T}\ev{\hat{H}(t)}{\Phi(t)}\dd{t},
	\end{equation}
	which becomes equivalent to the average energy $\bar{E}[\Phi]$ when the quasi-energy $\epsilon[\Phi]$ is stationary
	\begin{gather}
		\bar{\mathcal{E}}[\Phi]=\bar{E}[\Phi]\quad\Leftrightarrow\quad\Phi\in\qty{\Phi'\in\mathds{H}\otimes\mathds{T}\Big\vert\delta\epsilon[\Phi']=0},\\
		\epsilon[\Phi]=\frac{1}{T}\int_{0}^{T}\ev{\hat{H}(t)-i\hat{\partial}_t}{\Phi(t)}\dd{t}.
	\end{gather}
	So, in practice, the efficient Lagrangian minimization method becomes:
	\begin{equation}
		\bar{E}_0=\min_{\Phi\in\mathds{H}\otimes\mathds{T}}\qty\Big{\bar{\mathcal{E}}[\Phi]+\sum_i\lambda_i\fdv{\epsilon[\Phi]}{\Phi_i}+\mu\qty\Big(\ip{\Phi}-1)}, \label{Eq:Def min}
	\end{equation}
	where $\qty{\Phi_i}$ is an arbitrary basis spanning the Floquet space $\mathds{H}\otimes\mathds{T}$. We can thus straightforwardly calculate the ground state by constraint minimization methods in the Floquet space $\mathds{H}\otimes\mathds{T}$ using only easily calculable functionals. This offers an alternative to the conventional perturbation and adiabatic continuation method, and it does not require any approximation to the Floquet Hamiltonian $\hat{H}^F$, just adequate parametrization of the Floquet wave function.
	
	Secondly, the average energy is robust against infinitesimal perturbations. However, this was already mentioned in \cref{Sec:Difference}. Here we will add that, with the lower-bounded and unconfined ordering of the eigenstates $\Phi_{na}$ by the average energy $\bar{E}_{na}$, we can now systematically truncate the Floquet space $\mathds{H}\otimes\mathds{T}$ to a computationally accessible subspace, focusing on the lowest average energy states. This method would not require any adiabatic continuation to figure out which Floquet states are significant, or the prior evaluations of the static energy eigenstates, and is thus more computationally efficient and more generally applicable.
	
	Finally, as for the physical significance of the average energy, we expect that it can serve a similar role to the static energies in the thermal equilibrium. There have already been attempts at quantifying the Floquet steady state by the average energy~\cite{Ketzmerick_2010}, although using a different definition for the average energy, showing how we cannot generally formulate an equivalent Boltzmann distribution around it. This is mostly due to the steady state being defined primarily by the energy spectra, and its dependence on the detailed system-bath interaction. So instead we should consider the average energy as an approximation tool for systematically truncating the Hilbert space to a finite computable one. We can at the very least eliminate the highly excited states as long as the energy spectra are weakly overlapping~\cite{Le_Inpreparation}.
	
	\section{Conclusion\label{Sec:Conclusion}}
	To conclude, we report that a more complete method for studying the time-periodic quantum systems using Floquet theory is to decompose the system onto the Floquet eigenstates defined as the eigentriplet with the quasi-energy and average energy. One of the main property of this formulation is the introduction of the Ritz variational principle. The lack of this has been a major roadblock in the development of efficient Floquet first-principles methods, most beautifully exemplified in the incompatibility of the Floquet Density Functional Theory~\cite{Maitra_2002,Maitra_2007}. Now, we can revisit these Floquet first-principles methods, which would bring us one step closer to efficiently simulating the steady state of a many-body electron system under constant laser irradiation. Indeed, the Floquet Hartree-Fock method is straightforward to derive~\cite{Le_Inpreparation}.

	The main physical significance of the average energy would be in calculating the steady state. We do not expect it to have exactly the same role as the static energy, but instead it should have a supporting one, such as to limit the Hilbert space to the physically significant states. To that end, the average energy ground state might not have a direct physical significance, but we conjecture that it generally approximates the steady state with dense thermal baths. Indeed we can check the validity of these statements in various model systems, while a quantitative proof is still being developed.
	
	This work was supported by JST-Mirai Program Grant Number JPMJMI20A1, Japan. C. M. Le was supported by the Japan Society for the Promotion of Science through the Program for Leading Graduate Schools (MERIT).

	\appendix
	\section{Proof of $\comm*{\hat{\bar{H}}}{\hat{T}_S}=0$\label{App:Comm}}
	We can rewrite the operators $\hat{\bar{H}}$ and $\hat{T}_S$ as
	\begin{gather}
		\hat{\bar{H}}=\sum_{nij}\bar{H}_{nij}\op{\Psi'_{ni}}{\Psi'_{nj}}, \label{Eq:App:barH}\\
		\hat{T}_S=\sum_{ni}e^{-i\epsilon_nT}\op{\Psi'_{ni}}. \label{Eq:App:Ts}
	\end{gather}
	Using an arbitrary quasi-energy eigenbasis $\qty{\Psi'_{ni}}$ (\cref{Eq:Phini}), for which we will reserve the primed notation here for clarity. This basis satisfies
	\begin{gather}
		\ket{\Psi'_{ni}}=\lim\limits_{\mathcal{T}\to\infty}\frac{1}{\sqrt{2\mathcal{T}}}\int_{-\mathcal{T}}^{\mathcal{T}}e^{-i\epsilon_nt}\ket{\Phi'_{ni}(t)}\otimes\ket{t}\dd{t},\\
		\qty[\hat{H}(t)-i\partial_t]\ket{\Phi'_{ni}(t)}=\epsilon_{n}\ket{\Phi'_{ni}(t)},
	\end{gather}
	so that we retain the orthonormality conditions
	\begin{equation}
		\ip{\Psi'_{mi}}{\Psi'_{nj}}=\delta_{mn}\delta_{ij}.
	\end{equation}
	Combining this orthonormality condition with the definitions in \cref{Eq:App:barH,Eq:App:Ts} we straightforwardly get:
	\begin{widetext}
		\begin{align}
			\comm{\hat{\bar{H}}}{\hat{T}_S}&=\sum_{mnijk}\qty[\bar{H}_{mij}e^{-i\epsilon_nT}\delta_{mn}\delta_{jk}-e^{-i\epsilon_mT}\bar{H}_{njk}\delta_{mn}\delta_{ij}]\op{\Psi'_{mi}}{\Psi'_{nk}}, \\
			&=\sum_{nij}\qty[\bar{H}_{nik}e^{-i\epsilon_nT}-e^{-i\epsilon_nT}\bar{H}_{nik}]\op{\Psi'_{ni}}{\Psi'_{nk}}=0.
		\end{align}
	\end{widetext}

	The commutation of the effective operator $\hat{\bar{H}}_n$ and the Floquet Hamiltonian $\hat{H}^F$ in the Floquet space follows the same steps.

	\section{Derivation of $\bar{H}_n$\label{App:Hn}}
	Starting from the physical eigenstate definition,
	\begin{gather}
		\hat{\bar{H}}\ket{\Psi_{na}}=\bar{E}_{na}\ket{\Psi_{na}}, \label{Eq:App:HPsi}\\
		\ket{\Psi_{na}}=\lim\limits_{\mathcal{T}\to\infty}\frac{1}{\sqrt{2\mathcal{T}}}\int_{-\mathcal{T}}^{\mathcal{T}}e^{-i\epsilon_nt}\ket{\Phi_{na}(t)}\otimes\ket{t}\dd{t},
	\end{gather}
	we operate $\bra{t}$ on the left of \cref{Eq:App:HPsi}, where $t$ is an arbitrary time:
	\begin{equation}
		\mel{t}{\hat{\bar{H}}}{\Psi_{na}}=\bar{E}_{na}\ip{t}{\Psi_{na}},
	\end{equation}
	We expand the definition of $\hat{\bar{H}}$ from \cref{Eq:App:barH}
	\begin{equation}
		\sum_{mij}\bar{H}_{mij}\ip{t}{\Psi'_{mi}}\ip{\Psi'_{mj}}{\Psi_{na}}=\bar{E}_{na}\ip{t}{\Psi_{na}}, \label{Eq:App:HmijPsi}
	\end{equation}
	The inner product of the state vectors is simplified as
	\begin{align}
		\ip{\Psi'_{mj}}{\Psi_{na}}&=\lim\limits_{\mathcal{T}\to\infty}\frac{1}{2\mathcal{T}}\int\limits_{-\mathcal{T}}^{\mathcal{T}}e^{i(\epsilon_m-\epsilon_n)t}\ip{\Phi_{mi}(t)}{\Phi_{na}(t)}\dd{t},\\
		&=\delta_{mn}\frac{1}{T}\int_{0}^{T}\ip{\Phi_{mi}(t)}{\Phi_{na}(t)}\dd{t}. \label{Eq:App:IP2}
	\end{align}

	Now it is helpful to introduce a double bra-ket notation for the vectors in the Floquet space $\mathds{H}\otimes\mathds{T}$, such that the inner-product is defined by:
	\begin{gather}
		\iipp{\Phi}{\Phi'}=\frac{1}{T}\int_{0}^{T}\ip{\Phi(t)}{\Phi'(t)}\dd{t}\qquad\forall\Phi,\Phi'\in\mathds{H}\otimes\mathds{T}. \label{Eq:App:Def iipp}
	\end{gather}
	Using \cref{Eq:App:Def iipp,Eq:App:IP2} in \cref{Eq:App:HmijPsi} and re-arranging some terms, we get:
	\begin{equation}
		\sum_{ij}\bar{H}_{nij}\ket{\Phi'_{ni}(t)}\iipp{\Phi'_{nj}}{\Phi_{na}}=\bar{E}_{na}\ket{\Phi_{na}(t)}.
	\end{equation}
	As this relation holds at all times $t$, we can expand it to the whole Floquet space $\mathds{T}$:
	\begin{equation}
		\qty\Big[\sum_{ij}\bar{H}_{nij}\oopp{\Phi'_{ni}}{\Phi'_{nj}}]\kket{\Phi_{na}}=\bar{E}_{na}\kket{\Phi_{na}}.
	\end{equation}
	Defining the bracketed expression as $\hat{\bar{H}}_n$, we get the original equation in \cref{Eq:Floq Eig1}. Repeating this process for all physical eigenstates $\Psi_{na}$ and for all quasi-energy Brillouin zones, we get the complete definition of $\hat{\bar{H}}_n$ spanning all Floquet space $\mathds{H}\otimes\mathds{T}$.

	\section{Equivalence of $\bar{\mathcal{E}}$ and $\bar{E}$\label{App:Equiv}}
	Expanding the functionals $\bar{E}[\Phi]$ and $\bar{\mathcal{E}}[\Phi]$ on an arbitrary quasi-energy basis $\qty{\Phi'_{ni}}$, we find the differences between these to be on the non-degenerate components with respect to the quasi-energy labels, i.e.:
	\begin{align}
		\bar{E}[\Phi]&=\sum_{nij}C^*_{ni}C_{nj}\bar{H}_{nij}, \label{Eq:App:AE Decomp}\\
		\bar{\mathcal{E}}[\Phi]&=\sum_{mnij}C^*_{mi}C_{nj}\bar{H}_{mnij}, \label{Eq:App:EAE Decomp}
	\end{align}
	where $\bar{H}_{mnij}$ is the more general uncontracted form of $\bar{H}_{nij}$ (\cref{Eq:Hnij}), representing the matrix element of the Hamiltonian:
	\begin{equation}
		\bar{H}_{mnij}=\frac{1}{T}\int_{0}^{T}\mel{\Phi'_{mi}(t)}{\hat{H}(t)}{\Phi'_{nj}(t)}\dd{t},
	\end{equation}
	and $C_{nj}$ is the usual overlap matrix with the arbitrary Floquet state $\Phi$:
	\begin{equation}
		C_{nj}=\frac{1}{T}\int_{0}^{T}\ip{\Phi'_{nj}(t)}{\Phi(t)}\dd{t}.
	\end{equation}
	So the condition for the two functionals to be equal simplifies to
	\begin{equation}
		\bar{E}[\Phi]=\bar{\mathcal{E}}[\Phi]\quad\Leftrightarrow\quad C^*_{mi}C_{nj}=\delta_{mn}C^*_{ni}C_{nj}. \label{Eq:App:Equiv cond}
	\end{equation}

	This condition is satisfied by the quasi-energy variation $\delta\epsilon[\Phi]=0$, which we will show explicitly here. For this we decompose the Floquet state $\Phi$ using two parameters $\theta,\varphi$ so that we focus on the two states $\Phi'_{mi}$ and $\Phi'_{nj}$ as follows:
	\begin{align}
		\ket{\Phi}&=\cos(\theta)\cos(\varphi)\ket{\Phi'_{mi}}+\cos(\theta)\sin(\varphi)\ket{\Phi'_{nj}}\notag\\
		&\phantom{=}+\sin(\theta)\ket{\Phi^{\prime\perp}}.
	\end{align}
	Here $\Phi^{\prime\perp}$ is the remaining projection of the Floquet state $\Phi$, orthogonal to both $\Phi'_{mi}$ and $\Phi'_{nj}$. We will also ignore the complex phase as it does not affect the quasi-energy variation. As the variation $\delta\epsilon[\Phi]=0$ has to hold for all parameters, the following condition has to hold
	\begin{equation}
		\fdv{\epsilon[\Phi]}{\varphi}=-(\epsilon_m-\epsilon_n)\cos^2(\theta)\sin(2\varphi)=0
	\end{equation}
	So the necessary condition for the quasi-energy variation to hold is:
	\begin{equation}
		\delta\epsilon[\Phi]=0\;\Rightarrow\;\left[\begin{aligned}
			&\epsilon_m=\epsilon_n&\Leftrightarrow&m=n;\quad\forall C_{mi},C_{nj},\\
			&\sin(2\varphi)=0&\Leftrightarrow&C_{mi}=0\;\mathrm{or}\;C_{nj}=0,\\
			&\cos(\theta)=0&\Leftrightarrow&C_{mi}=C_{nj}=0.
		\end{aligned}\right.
	\end{equation}
	Applying this for all pairs we get the equivalence with \cref{Eq:App:Equiv cond}:
	\begin{equation}
		\delta\epsilon[\Phi]=0\quad\Leftrightarrow\quad C^*_{mi}C_{nj}=\delta_{mn}C^*_{ni}C_{nj}.
	\end{equation}

	\bibliography{References.bib}

\end{document}